\documentclass[conference]{IEEEtran}

\ifCLASSINFOpdf
\usepackage[pdftex]{graphicx}
\usepackage{graphicx}
\else
\fi
\usepackage{graphicx}
\usepackage{wrapfig}
\usepackage{balance}
\usepackage{cite}
\usepackage{float}
\usepackage{amsmath}
\usepackage[utf8]{inputenc}
\usepackage[final]{pdfpages}
\usepackage{pdfpages}
\usepackage{lipsum}
\usepackage{textcase}
\usepackage{url}
\usepackage{amsmath,esint}
\usepackage{epstopdf}
\usepackage{array}
\usepackage{color}
\usepackage{pgfpages}
\pgfpagesuselayout{resize to}[a4paper]
\usepackage{subcaption}
\usepackage{multicol}

\usepackage{bm}

\DeclareRobustCommand{\uvec}[1]{{%
  \ifcsname uvec#1\endcsname
     \csname uvec#1\endcsname
   \else
    \bm{\hat{\mathbf{#1}}}%
   \fi
}}

\newcolumntype{P}[1]{>{\centering\arraybackslash}p{#1}}
\newcolumntype{M}[1]{>{\centering\arraybackslash}m{#1}}

\begin{document}
	
\title{Improving LoRa Signal Coverage in Urban and Sub-Urban Environments with UAVs
}

\author{
\IEEEauthorblockN{Vageesh Anand Dambal$^{1}$, Sameer Mohadikar$^{1}$, Abhaykumar Kumbhar$^{2}$ and Ismail Guvenc$^{1}$ 
}
\IEEEauthorblockA{$^1$Department of Electrical and Computer Engineering, North Carolina State University, Raleigh, NC \\
                  $^2$Department of Electrical and Computer Engineering, Florida International University, Miami, FL}
Email: \{vadambal, smohadi, iguvenc\}@ncsu.edu, akumb004@fiu.edu	
}

\maketitle


\begin{abstract}
LoRa technology enables long-range communication with low-power consumption for the Internet-of-Things (IoT) devices in the urban and suburban environment. However, due to terrestrial structures in urban and suburban environments, the link distance of LoRa transmissions can be reduced. In this paper, we report signal strength measurements for the in-building and inter-building LoRa links and provide insights on factors that affect signal quality such as the spreading factor and antenna orientation. Subsequently, we also provide measurement results in urban and suburban environments when the LoRa transmitter is deployed at different heights using an unmanned aerial vehicle (UAV). Our findings show that the UAV deployment height is critical for improving coverage in the suburban environment and antenna orientation affects the communication range.   



\begin{IEEEkeywords}
Antenna orientation, drone, indoor, IoT, LoRa, LP-WAN, suburban, UAV, urban.
\end{IEEEkeywords}

\end{abstract}

\IEEEpeerreviewmaketitle


    

\section{Introduction}

LoRa is a wireless technology, which is employed for achieving a long-range communication with diverse Internet of things (IoT) applications distributed over a wide geographical area. Some of the LoRa-based IoT applications include smart meter applications, infrastructure monitoring, smart well-being applications, vehicular tracking, and industrial monitoring and control~\cite{j1,j2,j3}. LoRa uses chirp spread spectrum (CSS) modulation scheme, wherein the carrier frequency decreases or increases over a specific amount of time thus achieving low power and long-range communication links~\cite{j1,j2}.

\begin{figure}[t]
	\centering\vspace{-4mm}
    \includegraphics[width=6.75cm]{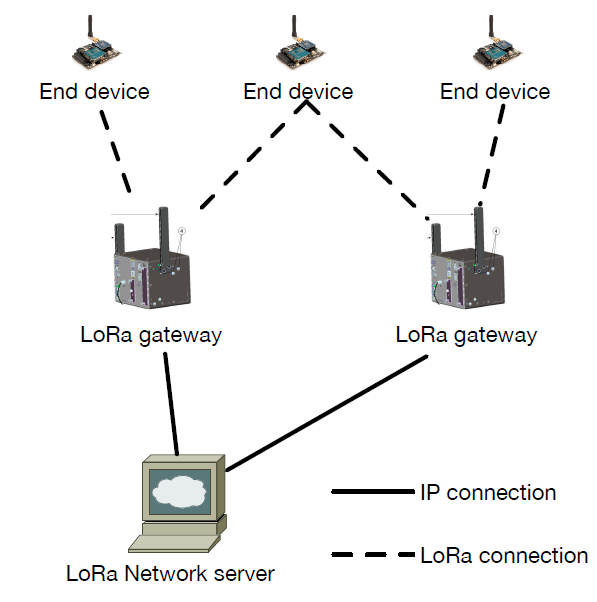}\vspace{-1mm}
    \caption{LoRa network architecture comprising of LoRa server, gateway, and end devices~\cite{c2}.}
    \label{fig:arch}
    \vspace{-3mm}
\end{figure}

A typical LoRa network is “a star-of-stars topology,” which includes three different types of nodes, i.e., a  LoRa server, LoRa gateway (GW), and end devices as illustrated in Fig.~\ref{fig:arch}~\cite{c2}. A single LoRa GW can cover an entire city or hundreds of square kilometers. However, this communication range is highly dependent on the environmental factors and terrestrial obstructions at a given location. Therefore, it would require complex network management algorithms and deployment process to build a scalable LoRa network with a ubiquitous communication link that supports millions of IoT devices in a given area. Furthermore, a LoRa GW attached on an aerial platform such as an unmanned aerial vehicle (UAV) can increase the existing coverage in a given area and thus complementing the existing infrastructure. The air-to-ground link quality can be further improved with the help of three dimensional (3D) radiation pattern of the antenna attached to LoRa GW~\cite{c1}.



Several of the recent works have focused on testing the range of the LoRa technology in different scenarios~\cite{c2,c3,c4}. Whereas, in literature~\cite{c5} a UAV based sensor network is proposed for a marine environment without considering the LoRa technology. To our best knowledge, LoRa link budget experiments with aerial platforms for an urban environment have not been reported. In this paper, we measure the received signal strength (RSS) at a LoRa GW in three different urban/suburban scenarios. In \textit{scenario-1}, both the LoRa GW and the end node are placed indoors,  whereas in \textit{scenario-2} the LoRa GW is placed outdoors in a suburban environment and the end node is placed indoors. Finally, in \textit{scenario-3}, a UAV is used to increase the LoRa range in an urban environment. 

The rest of this paper is organized as follows. In Section~\ref{sec:expsetUp} discusses the type of hardware, antenna configurations, and location used in the experimental setup. In Section~\ref{indoorExp}, we present our measurements for indoor experiments and outdoor experiment in Section~\ref{outExp}. Finally, the last section provides concluding remarks.

\section{Experiment Setup}
\label{sec:expsetUp}



In this section, we describes our measurement setup using the \textit{LoRa Technology Evaluation Kit–900} by Microchip Technology which has a bandwidth of $125$ kHz and operating between $902-928$ MHz. 
Our experimental LoRa network topology is similar to the network architecture illustrated in Fig.~\ref{fig:arch}. The RN2903 Mote boards are the end devices which communicate with the 8-channel gateway. The mote has a receiver sensitivity of $-123$~dBm and transmit power of $6$~dBm which is adjustable to $18.5$~dBm. The LoRa gateway core board acts as the gateway, which is responsible for forwarding data to single LoRa network server over Internet~\cite{c8}.

\begin{figure}
\centering\vspace{-3mm}
    \includegraphics[width=8.5cm]{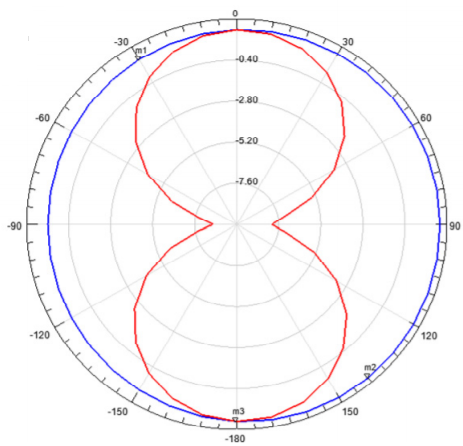} \vspace{-1mm}
    \caption{LoRa antenna radiation pattern. The red plot (similar to ''8'') is in the elevation plane while the blue plot is in the azimuth plane~\cite{c8}.}
    \label{rad}
    \vspace{-3mm}
\end{figure}

The LoRa protocol has an adjustable data rate, which varies with the spreading factor (SF). A larger SF implies a higher receiver sensitivity, lower bit rate, and longer range. The horizontal and vertical radiation pattern of the LoRa antenna (PCB Trace antenna) given by the manufacturer is shown in Fig.~\ref{rad}, which will be taken into account while interpreting our measurements ~\cite{c8}. The theoretical coverage of LoRa is $15$~km for suburban and $5$~km for urban areas.

Our indoor experiments were carried out at NCSU Centennial Campus buildings, where the LoRa mote was placed on a desk about 1.5~m high on the 3rd floor of a building, and we were roaming on each floor with the GW in our hand to collect measurement data. Both the gateway and the mote were stationary at the time of transmission. For outdoor experiments, we used a mote attached to the DJI Phantom 4 Pro flying at the 25~m and 50~m height, while the GW was on the ground at roughly 1~m height.

\section{Indoor Ground-to-Ground Experiments}
\label{indoorExp}


\begin{figure}
\centering 
\begin{subfigure}[b]{0.47\textwidth}
   \includegraphics[width=1\linewidth]{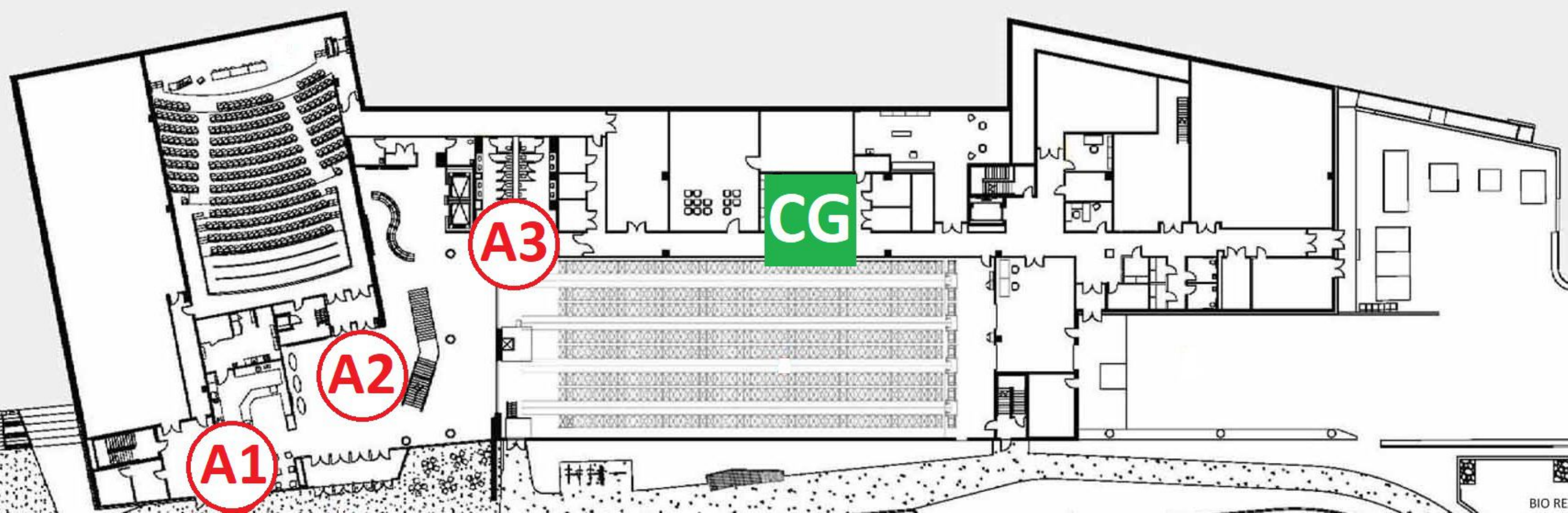}
   \caption{Floor-1 map of Hunt Library.} 
\end{subfigure}\vspace{4mm}
\begin{subfigure}[b]{0.47\textwidth}
   \includegraphics[width=1\linewidth]{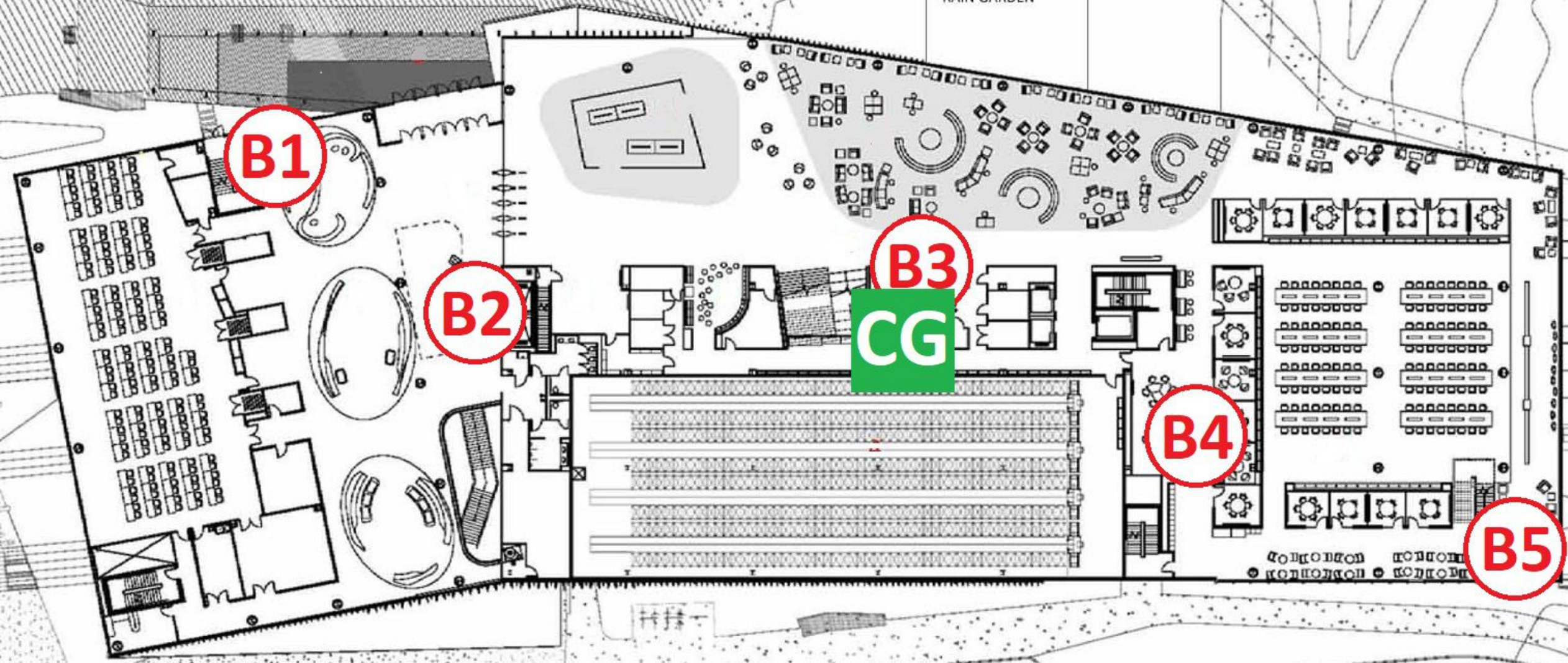}
   \caption{Floor-2 map of Hunt Library.}
\end{subfigure}\vspace{4mm}
\begin{subfigure}[b]{0.47\textwidth}
   \includegraphics[width=1\linewidth]{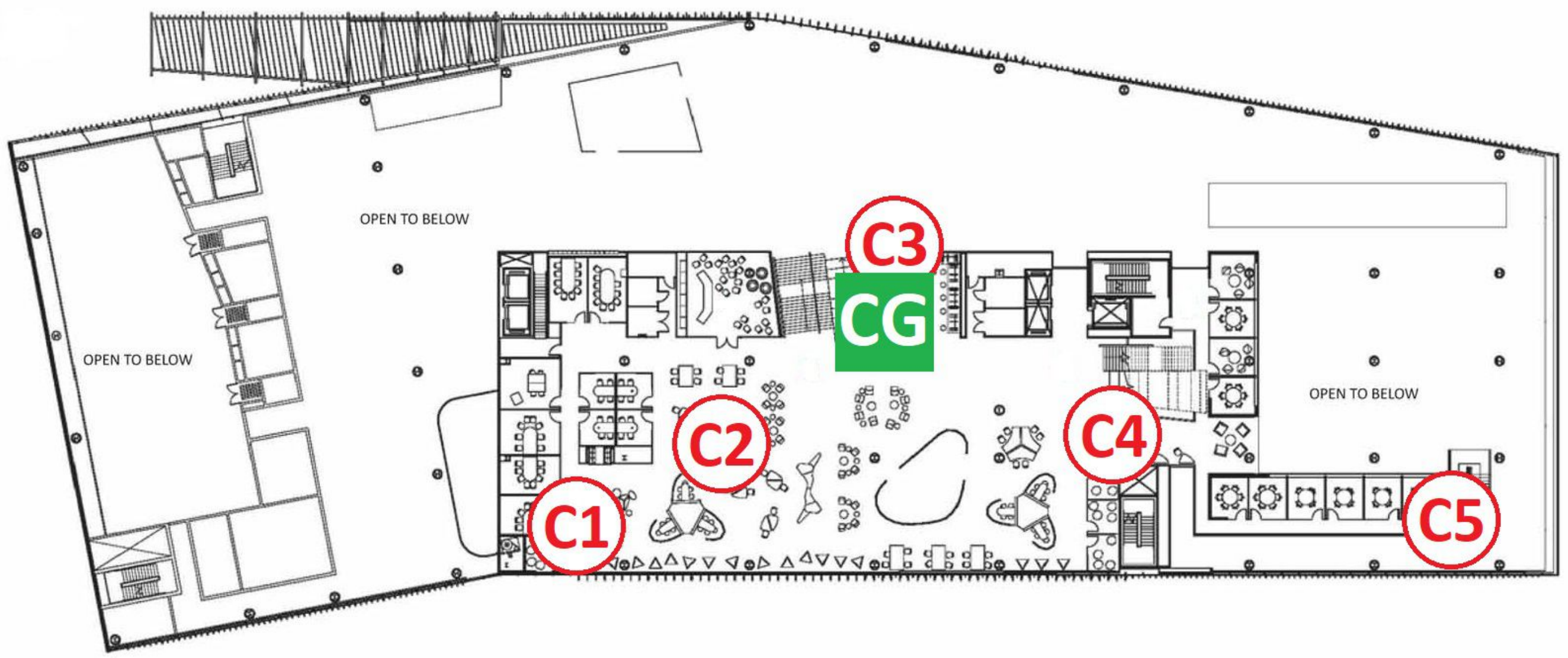}
   \caption{Floor-3 map of Hunt Library.} 
\end{subfigure}\vspace{4mm}
\begin{subfigure}[b]{0.47\textwidth}
   \includegraphics[width=1\linewidth]{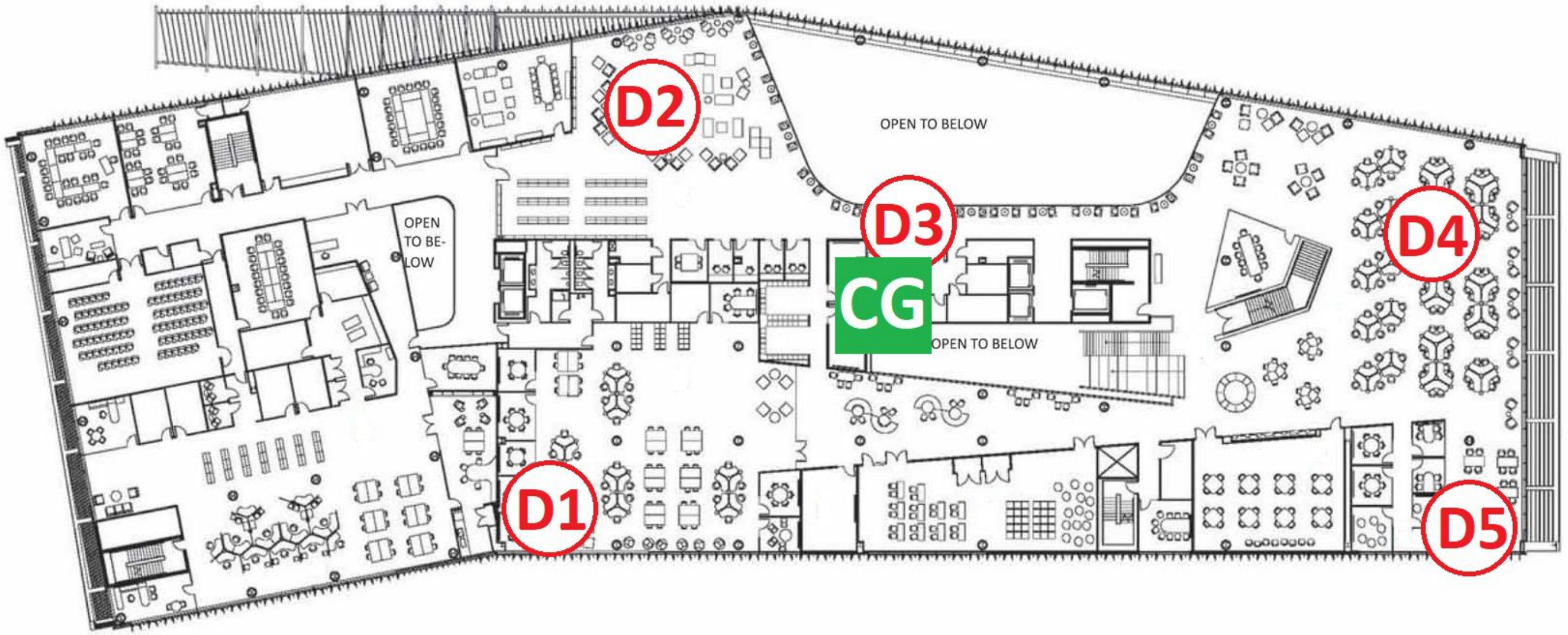}
   \caption{Floor-4 map of Hunt Library.}
\end{subfigure}\vspace{4mm}
\begin{subfigure}[b]{0.47\textwidth}
   \includegraphics[width=1\linewidth]{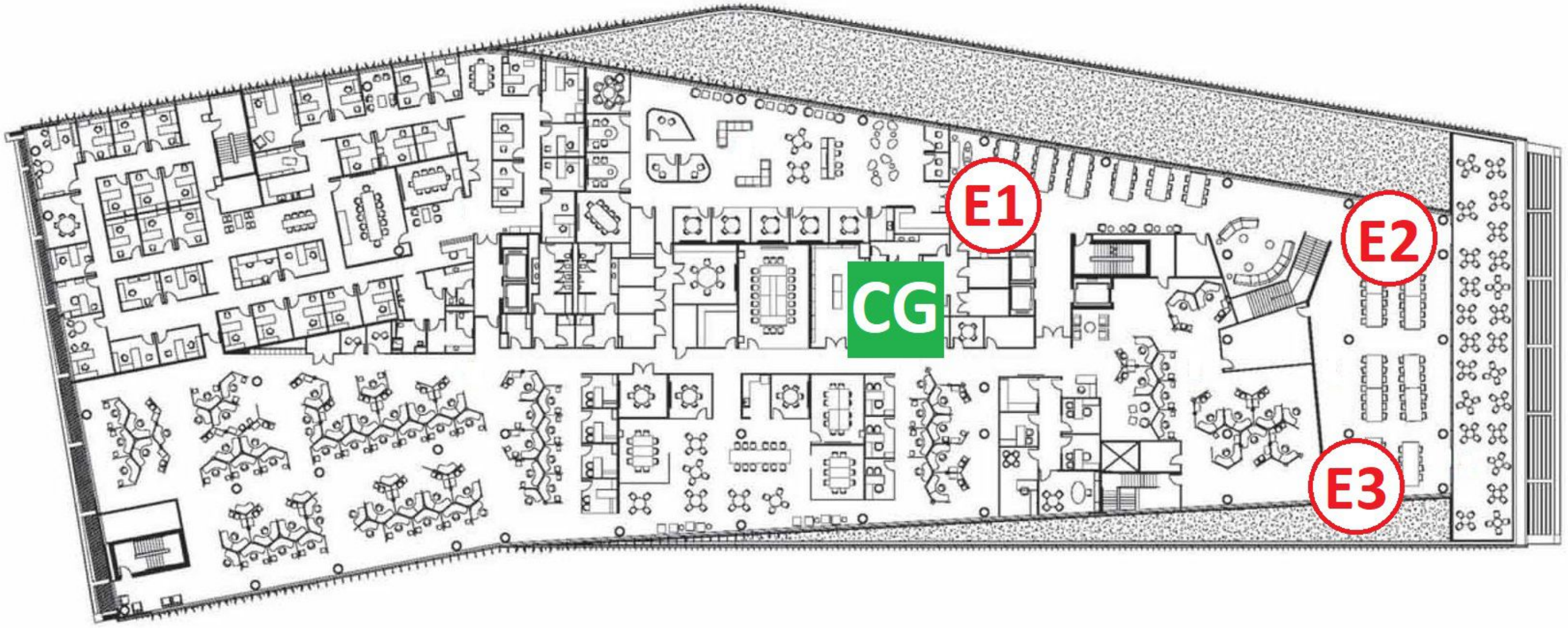}
   \caption{Floor-5 map of Hunt Library.}
\end{subfigure}

\caption{Position of LoRa GW/motes at NCSU Hunt Library. 'CG' denotes the location of the LoRa gateway that is placed permanently on the third floor of the library.}\label{diff_floors}
\vspace{-3 mm}
\end{figure}

\subsection{Indoor Transmission within a Building}

For the indoor setup, the LoRa GW is positioned on the third floor (out of five) of the Hunt Library in NCSU Centennial Campus, and the mote position is varied among different floors as shown in Fig.~\ref{diff_floors}, where CG refers to the fixed GW location on the third floor. Table~\ref{Tab1} shows the RSS at each measurement location for a spreading factor of 7 (SF7). Since a higher SF implies a better receiver sensitivity, if a packet is received with a SF7 then it is bound to be received with a SF greater than 7. However, this is true only if the GW and the node are within the same building as is shown in our experiment. The average RSS is shown in Table~\ref{Tab2} corroborates that even though the interior of a building may be void of any obstructions, the RSS is highly dependent on the floor where the GW is positioned.

\begin{table}[h!]
\centering \scriptsize
 \caption{RSS (in dBm) at different node positions.}\label{Tab1}
 \begin{tabular}{|c|c|c|c|c|c|c|c|c|c|} 
 \hline  {\bf Flr1} & {\bf RSS} & {\bf Flr2} & {\bf RSS} & {\bf Flr3} & {\bf RSS} & {\bf Flr4} & {\bf RSS} & {\bf Flr5} & {\bf RSS}  \\ 
 \hline  $\bf A1$ & $\bf -103$ & $\bf B1$ & $\bf -68$ & $\bf C1$ & $\bf -59$ & $\bf D1$ & $\bf -89$ & $\bf E1$ & $\bf -89$  \\  
 \hline  $\bf A2$ & $\bf -107$ & $\bf B2$ & $\bf -79$  & $\bf C2$ & $\bf -45$ & $\bf D2$ & $\bf -71$ & $\bf E2$ & $\bf -91$  \\  
 \hline  $\bf A3$ & $\bf -101$ & $\bf B3$ & $\bf -54$ & $\bf C3$ & $\bf -3$ & $\bf D3$ & $\bf -61$ & $\bf E3$ & $\bf -96$ \\ 
 \hline  - & - & $\bf B4$ & $\bf -65$  & $\bf C4$ & $\bf -71$ & $\bf D4$ & $\bf -73$ & - & - \\  
 \hline  - & - & $\bf B5$ & $\bf -79$ & $\bf C5$ & $\bf -88$ & $\bf D5$ & $\bf -72$ & - & -  \\ 
\hline
 \end{tabular}
 
\end{table}

As expected, the strongest RSS is observed at the third floor where the GW is positioned (and strongest at location C3 which is next to the GW), and two floors may introduce attenuation in the order of $50$~dB. The average RSS at the GW from different floors is further summarized in Table~\ref{Tab2}. Several interesting observations can be made based on these results; for example, even though A1 is further from the CG, it observes better RSS compared to A2, potentially due to outdoor reflections from large windows where A1 is located at.


\begin{table}
\centering
\vspace{4 mm}
 \caption{Average RSS (in dBm) at different node positions.}\label{Tab2}
 \begin{tabular}{|c|c||c|c|} 
 \hline {\bf Intra-building} & {\bf RSS} & {\bf Inter-building} & {\bf RSS} \\
 \hline\hline  $\bf Floor~1$ & $ -103.66$ & $\bf Pos~1$ & $ -99.5$ \\ 
 \hline  $\bf Floor~2$ & $ -69$ & $\bf Pos~2$ & $ -102$ \\ 
 \hline  $\bf Floor~3$ & $ -53.2$ & $\bf Pos~3$ & $ -102.75$\\ 
  \hline  $\bf Floor~4$ & $ -73.2$ & - & - \\ 
   \hline  $\bf Floor~5$ & $ -92$ & - & -\\ 
\hline
 \end{tabular}
 \vspace{-3 mm}
\end{table}

\subsection{Indoor Transmission between Buildings}

When the mote and the GW are in different buildings, the transmitted signal must penetrate two layers of wall/glass as a result of which we need a higher receiver sensitivity at the GW. To get a higher receiver sensitivity, we need to lower the data rate and increase the SF. We performed the experiment with a SF of 10 for LoRa transmission/reception between two different buildings. A lower SF would require the GW to have a lower receiver sensitivity, and hence we found that no reception is possible with an SF greater than 10 when the node is placed indoor in another building.

\begin{figure}
	\centering
    \includegraphics[width=5.5cm]{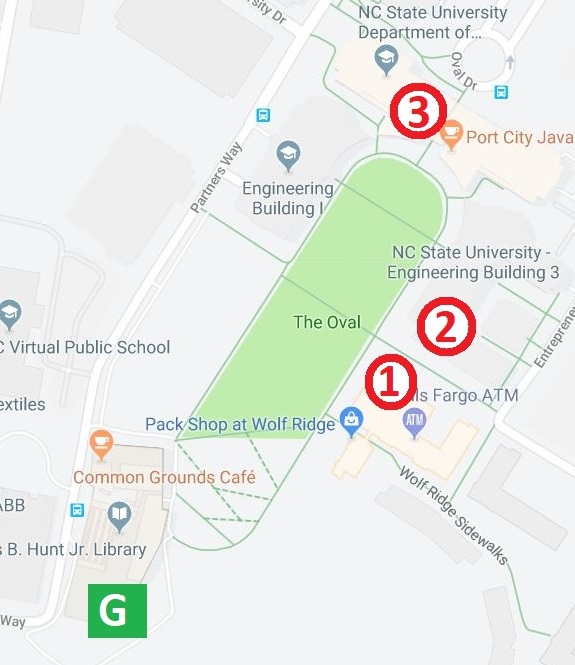}
    \caption{GW and mote locations for inter-building RSS measurements at NCSU Centennial Campus.}\label{intra_indoor}
    \vspace{-3 mm}
\end{figure}

We conducted experiments for the GW and mote locations illustrated in Fig.~\ref{intra_indoor} which shows the map of NCSU Centennial Campus. The GW is located on the third floor (indoors) of Hunt Library, while the motes are placed in three other buildings as specified in the figure. The distance between each building is approximately 150 m. The RSSs at the GW from these three different locations are captured in Table~\ref{Tab2}. 
Comparing the RSS for inter-building measurements with those of intra-building measurements in Table~\ref{Tab2}, we can see that link quality is severely degraded with inter-building propagation. 
We also found that the reception of the signal, in this case, depends strongly on the position of the node within the building. A relatively strong signal is received only if the node is present in the vicinity of a door/window of the building.

\section{Outdoor Air-To-Ground Experiment}
\label{outExp}
\begin{figure}[h]
	\centering
    \includegraphics[width=8.75cm]{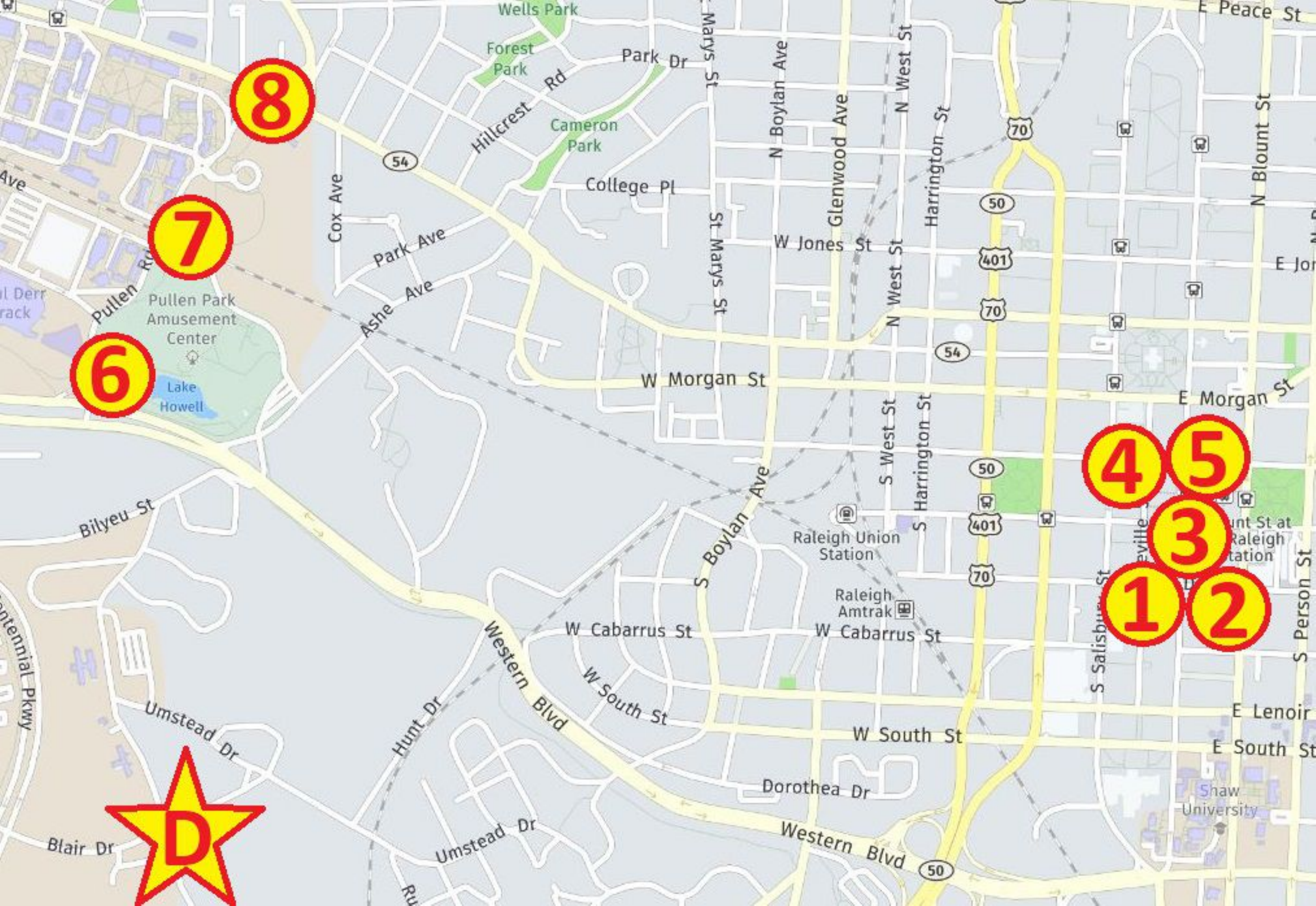}
    \caption{Measurement locations with UAV. Locations 1-5 are for downtown Raleigh,  locations 6-8 are for suburban areas.}\label{downtown_map}
    \vspace{-3 mm}
\end{figure}

After exploring signal propagation characteristics within/among buildings, we studied the signal propagation in long-distance urban environments and investigated whether UAVs can be used to improve coverage.  
The main goal of this experiment to measure the RSS when the GW is in suburban and urban areas with the mote mounted on a drone (DJI Phantom 4 Pro). The drone flies with the GW transmitter at heights of 25~m and 50~m, in a large park close to NCSU campus. While the receiver (GW) antenna was always vertically located, the transmitter (mote) antenna at the drone is held at both vertical and horizontal positions, corresponding to scenarios VV and VH, respectively. We covered both urban (locations 1-5, at Raleigh downtown with tall buildings) and suburban (locations 6-8) environments as shown in Fig.~\ref{downtown_map}. The approximate distance between the drone and the points in the suburban locations are 0.71 miles, 0.95 miles, and 1.12 miles. All the RSS measurements for these locations are presented in the tabular form in Table~\ref{Tab4}. 

From the antenna radiation pattern in Fig.~\ref{rad} we can conclude that when the antenna is perpendicular to the ground, the radiation pattern corresponds to the blue (outer circle). However, when the antenna is in the horizontal position, the radiation pattern corresponds to the inner red pattern which results in a lower antenna gain~[6]. Hence, for all VH scenarios (transmitter antenna at drone horizontal to the ground),  there is a significant decrease in the RSS compared to VH scenarios as can be observed from Table~\ref{Tab4}.

The last three readings in Table~\ref{Tab4} summarize our findings of the RSS in a suburban area. Since a suburban area does not contain high-rise buildings, there are no multi-path signals to impinge on the receiver, and hence the RSS decreases as the distance increases. We can also observe that the effect of drone height becomes more critical for signal reception in suburban areas when compared to urban environments. We can observe that the RSS at 50~m drone height is always better than the RSS at 25~m. We also attempted measurements when the transmitter is at 1~m height (rather than flying at the drone); however, no reception was observed for all the scenarios, implying that drones can be used to extend the coverage significantly.


\begin{table}[!h]
 \caption{RSS at various drone heights (1-5 Urban, 6-8 suburban). All values in dBm.}\label{Tab4}
\begin{center}
\begin{tabular}{| c || c | c || c | c |}
\hline
\textbf{Position}  &  \multicolumn{2}{ c |}{\textbf{Drone at 25 m Height.}}  &  \multicolumn{2}{ c |}{\textbf{Drone at 50 m Height.}}  \\ 
\cline{2-5}
& \textbf{V-V} & \textbf{V-H} & \textbf{V-V} & \textbf{V-H} \\
\hline

1 &  -116.6  & -117.28  & -115.92  & -117  \\ \hline
2 & -114.5  & -115.5  & -109.4  & -114.5 \\ \hline
3 & -119  & No Reception & -117.8  & No Reception\\ \hline
4 & -116.88  & -120  & -115.57  & -117 \\ \hline
5 & -115.13  & No Reception & -114.6  & -117 \\ \hline
6 & -118  & -- & -111  & -- \\ \hline
7 & -117  & --  & -113  & -- \\ \hline
8 & -118  & -- & -112  & -- \\ \hline

\end{tabular}
\vspace{-3 mm}
\end{center}
\end{table}

\section{Conclusion}

In this paper, we have measured the RSS at a LoRa GW for indoor, suburban, and urban areas; when the LoRa transmitter is in another indoor location or mounted on a UAV. For an indoor setting, we conclude that a LoRa GW can receive packets if placed in the direct line of sight of an opening in the building such as a door, window or a ledge.  
For the suburban environment, the drone height and antenna orientation play a crucial role on the RSS as there are no strong signal reflectors to cause multi-path reception at the LoRa GW. 

The drone heights at 25~m and 50~m do not have a significant effect on the RSS in urban areas as there exist a lot of multi-path signals that can impinge on the GW. Also if the transmitting antenna is vertical, a stronger signal is received due to the radiation pattern. 
The results also show that the LoRa network can operate effectively for up to 1.8 miles for air-to-ground transmission in urban areas. 

\section*{Acknowledgement}
This work has been supported in part by NASA under the grant number NNX17AJ94A, and by the National Science Foundation under the grant number CNS-1814727.

\end{document}